\documentclass[aps,pra,reprint,groupaddress]{revtex4-1}
\usepackage{lipsum}
\usepackage[utf8]{inputenc}
\usepackage[lmargin=25mm,rmargin=25mm,tmargin=27mm,bmargin=30mm]{geometry}
\usepackage{amsmath,bbm}
\usepackage{amssymb}
\usepackage{graphicx}
\usepackage{color}
\usepackage{url,hyperref}
\hypersetup{colorlinks=true,
	linkcolor=blue,
	citecolor=blue,
	urlcolor=blue,
	filecolor=blue
}

\newcommand{\bs}{\boldsymbol}
\newcommand{\ua}{\uparrow}
\newcommand{\da}{\downarrow}

\def\braket#1{\mathinner{\langle{#1}\rangle}}
\def\bra#1{\langle{#1}|}
\def\ket#1{|{#1}\rangle}

\begin{document}

\title{Transport-induced suppression of nuclear field fluctuations in multi-quantum-dot systems}

\author{Jørgen Holme Qvist}
\author{Jeroen Danon}%
\affiliation{Center for Quantum Spintronics, Department of Physics, Norwegian University of Science and Technology, NO-7491 Trondheim, Norway}

\date{\today} 

\begin{abstract}
Magnetic noise from randomly fluctuating nuclear spin ensembles is the dominating source of decoherence for many multi-quantum-dot multielectron spin qubits.
Here we investigate in detail the effect of a DC electric current on the coupled electron-nuclear spin dynamics in double and triple quantum dots tuned to the regime of Pauli spin blockade.
We consider both systems with and without significant spin-orbit coupling and find that in all cases the flow of electrons can induce a process of dynamical nuclear spin polarization that effectively suppresses the nuclear polarization gradients over neighboring dots.
Since exactly these gradients are the components of the nuclear fields that act harmfully in the qubit subspace, we believe that this presents a straightforward way to extend coherence times in multielectron spin qubits by at least one order of magnitude.

\end{abstract}

\maketitle

\section{Introduction}

Spin qubits hosted in semiconductor quantum dots form an attractive qubit implementation that promises easily scalable quantum processors~\cite{Hanson2007,Zwanenburg2013,Vandersypen2017}.
One drawback of the originally proposed single-spin single-quantum-dot qubit is that it requires highly localized magnetic fields for qubit control~\cite{Loss1997,Koppens2006}.
To overcome the practical challenge of creating such fields, qubits can also be encoded in a \emph{multielectron} spin state hosted in a multi-quantum-dot structure.
If one defines a qubit in the unpolarized singlet-triplet subspace of two spins in a double quantum dot, then the field along one axis of the Bloch sphere can be controlled fully electrically, but the second control axis is still set by the magnetic field gradient over the two dots~\cite{Petta2005,Taylor2007}.
Adding one more spin to the setup, one can create a three-electron double-dot hybrid qubit~\cite{Shi2012,Kim2014} or a triple-dot exchange-only qubit~\cite{Laird2010, Gaudreau2012, Medford2013, Medford2013_2, Taylor2013}, offering electric control over the full Bloch sphere through exchange interactions~\cite{DiVincenzo2000,russ2017three}.

An important remaining challenge for many multispin qubit implementations is their rapid decoherence.
Its two main sources are
(i) hyperfine coupling of the electronic spins to the randomly fluctuating nuclear spin baths in the quantum dots~\cite{Merkulov2002, Khaetskii2002, Hung2014, Peterfalvi2017}
and (ii) charge fluctuations in the environment that interfere with exchange-based qubit control~\cite{Hu2006,Russ2015}.
The latter could be mitigated by enhancing device quality or operating the qubit at a (higher-order) sweet spot \cite{Martins2016, Reed2016, Shim2016, Malinowski2017, Zhang2018}, which leaves the nuclear spin noise as an important intrinsic obstacle for further progress.

Several approaches to reducing the harmful effects of nuclear spin fluctuations in exchange-only qubits are being explored:
(i) One can host the qubits in quantum dots created in isotopically purified ${}^{28}$Si, which can be made nearly nuclear-spin-free~\cite{Muhonen2014, Enge1500214, purified_silicon1, purified_silicon3, Yoneda2018}.
However, silicon comes with the complication of the extra valley degree of freedom~\cite{Zwanenburg2013}, which is hard to control~\cite{Friesen2007,Culcer2010,Neyens2018} and provides an extra channel for leakage and dephasing~\cite{Tahan2014,Sala2018}.
(ii) It is possible to encode the qubit in a four-electron singlet-only subspace \cite{Sala2017, Russ2018, sala2019highly}, which makes it intrinsically insensitive to the fluctuating nuclear fields.
This, however, presents significant complications for device design and tuning.
(iii) One can actively mitigate the nuclear spin noise, e.g., by applying complex spin-echo-like pulse sequences that effectively filter out all peaks from the noise spectrum~\cite{Malinowski2016} or with an active feedback cycle that relies on continuous measurement of the magnitude of the nuclear fields~\cite{Bluhm2010}.

In this paper we propose another approach that falls in the last category but is much simpler to implement.
A few years ago, experiments on a double quantum dot hosted in an InAs nanowire suggested that when running a DC electric current through the system in the regime of Pauli spin blockade, an interplay between the hyperfine interaction and strong spin-orbit interaction (SOI) in InAs can give rise to a process of dynamical nuclear polarization that effectively quenches the total Zeeman gradient over the two dots~\cite{spin-blockade-quenching}.
Here, we investigate this idea in more detail, and we show how it not only works for double quantum dots with strong SOI, but also in the absence of SOI and---maybe more importantly---can be implemented in a similar way in a linear triple quantum dot, where it results in a suppression of \emph{both} nuclear field gradients between neighboring dots.
For all mechanisms we investigate, we present a simple intuitive picture as well as analytic and numerical results that support this picture and predict a suppression of the fluctuations of the nuclear field gradients of one to two orders of magnitude.
Since hyperfine-induced decoherence of both singlet-triplet and exchange-only qubits originates mainly from these gradients, we believe that this current-induced suppression mechanisms yields a straightforward way to significantly extend the coherence time of multielectron qubits. 

The rest of this paper is separated into two main parts, Secs.~\ref{sec:double_dot} and \ref{sec:triple_dot}, which discuss the double-dot and triple-dot setup, respectively.
Both parts are are organized as follows:
In Subsections A we briefly review the definition of the respective qubit and present a description of the system in terms of a simple model Hamiltonian.
In Subsections B we then present an intuitive picture of the mechanism behind the suppression of the gradients.
Subsections C contain approximate analytic expressions for the current-induced dynamics of the nuclear polarizations, which we corroborate in Subsections D with numerical simulations of the stochastic nuclear spin dynamics.
Subsections E contain a short conclusion, and a final general conclusion is presented in Sec.~\ref{sec:conclusion}.

\section{Singlet-triplet qubit}
\label{sec:double_dot}

\subsection{The qubit}

\begin{figure}
	\includegraphics[width=.43\textwidth]{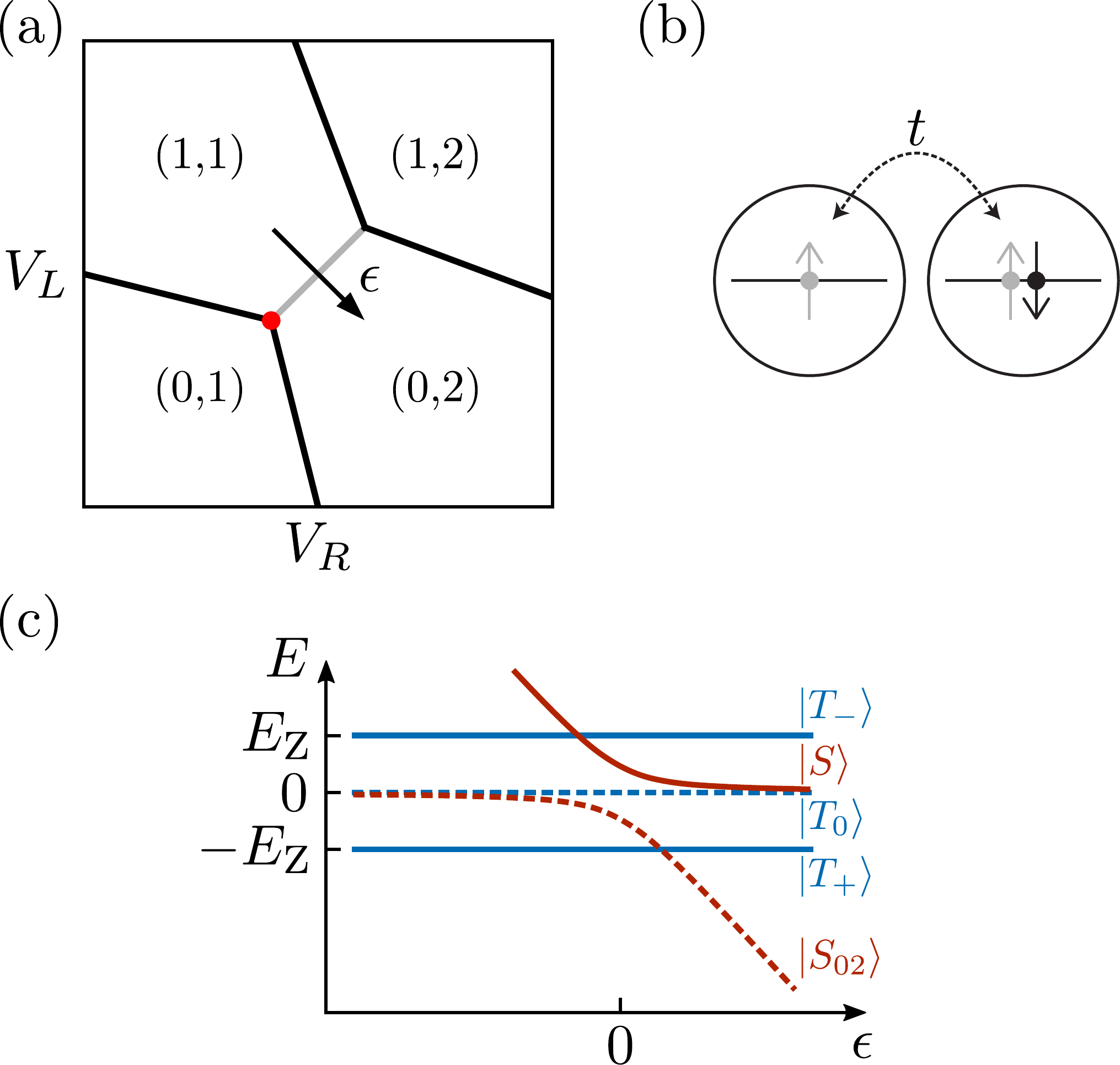}
	\caption{
		(a) Typical charge stability diagram of a double quantum dot, showing the ground state charge configuration of the system as a function of the local dot potentials $V_L$ and $V_R$.
		(b) Sketch of the double quantum dot.
		If the system is in a (0,2) charge state, the two electrons must have opposite spin.
		(c) Energy spectrum along the detuning axis indicated in (a) showing the relevant (1,1) and (0,2) states, where a finite interdot tunnel coupling and Zeeman splitting were included.
		The blue(red) lines correspond to spin triplet(singlet) states.}
	\label{fig:model}
\end{figure}

The singlet-triplet qubit is usually hosted by two electrons residing in a double quantum dot and is defined in two two-particle spin states with total spin projection $S_z = 0$.
Using gate voltages, the double dot is tuned 
close to the (1,1)--(0,2) charge transition (the gray line in the charge stability diagram shown in Fig.~\ref{fig:model}a).
Here, the low-energy part of the spectrum consists of five states:
The large orbital level splitting on the dots (typically $\sim$~meV) allows us to disregard states involving excited orbital states; the Pauli exclusion principle then dictates that the two electrons in the (0,2) configuration must be in a spin-singlet state, $\ket{S_{02}}$.
In the (1,1) charge configuration all four spin states are accessible; one singlet state $\ket{S}$, and three triplet states $\ket{T_\pm}$ and $\ket{T_0}$.

We describe this five-level subspace with a simple model Hamiltonian,
\begin{align}
\hat H_0 = \hat H_e + \hat H_t + \hat H_{\rm Z}.
\end{align}
Here
\begin{equation}
\hat{H}_e = -\epsilon \ket{S_{02}}\bra{S_{02}},
\end{equation}
describes the relative energy detuning of the (1,1) and (0,2) charge states as a function of the detuning parameter $\epsilon$, see Fig.~\ref{fig:model}.
Further,
\begin{align}
\hat{H}_t = t_s \big[ \ket{S}\bra{S_{02}}  + \ket{S_{02}}\bra{S} \big],
\end{align}
accounts for spin-conserving interdot tunneling, and
\begin{align}
\hat{H}_{\rm Z} = &
g\mu_{\rm B}B \big[ \ket{T_+}\bra{T_+}-\ket{T_-}\bra{T_-}\big]
,
\label{eq:Zeeman}
\end{align}
describes the Zeeman effect due to a homogeneous magnetic field.
A typical spectrum of $\hat H_0$ as a function of $\epsilon$ is shown in Fig.~\ref{fig:model}c, where we have set $t_s = 0.6\, E_{\rm Z}$ with $E_{\rm Z} = |g\mu_{\rm B}B|$ the Zeeman splitting, and we assumed $g<0$.

The qubit is defined in an unpolarized subspace consisting of a triplet, $\ket{1} = \ket{T_0}$, and the lower of the two singlet branches, $\ket{0} = \ket{S_2} = \cos \frac{\theta}{2} \ket{S_{02}} + \sin \frac{\theta}{2} \ket{S}$ (dashed levels in Fig.~\ref{fig:model}c) where $\tan \theta = 2t_s/\epsilon$.
From the projected qubit Hamiltonian
\begin{align}
\hat H_q = \frac{\omega_q}{2} \hat \sigma_z,
\end{align}
with $\omega_q = \epsilon/2 + \sqrt{(\epsilon/2)^2 + t_s^2}$ we see that the qubit has a splitting that is tunable electrically via $V_{L,R}$, presenting an advantage over the single-spin qubit, which requires magnetic control.

In semiconductors with non-zero nuclear spin, such as GaAs and InAs, an important source of decoherence for such a qubit is the hyperfine interaction between the nuclear and electronic spins.
The dominating term is the contact interaction, described by
\begin{equation}
\hat{H}_\text{hf}
= \frac{A}{2N}\sum_{d,k}\left(2\hat{S}^z_d\hat{I}^z_{d,k} + \hat{S}_d^+\hat{I}_{d,k}^- + \hat{S}_d^-\hat{I}_{d,k}^+\right),
\label{eq:hyperfine_hamiltonian}
\end{equation}
where $\hat {\bf S}_d$ is the electron spin operator on dot $d$ and $\hat {\bf I}_{d,k}$ the nuclear spin operator for nucleus $k$ on dot $d$.
For simplicity we assumed that all nuclei in a dot are coupled equally strongly to the electron spin in that dot and that both dots have the same number of spinful nuclei $N$, typically $N\sim 10^5$--$10^6$.
The coupling constant $A$ is a material parameter and usually of the order $\sim 100~\mu$eV.
Due to the small nuclear magnetic moment, the nuclear spin ensemble is in a fully mixed state in equilibrium at typical dilution fridge temperatures, and within a mean-field approximation we can then write
\begin{equation}
\hat{H}_\text{hf,mf}
= {\bf K}_L \cdot \hat {\bf S}_L + {\bf K}_R \cdot \hat {\bf S}_R,
\label{eq:hfmf}
\end{equation}
where the nuclear fields ${\bf K}_{L,R}$ are random with an r.m.s.\ value $\sim A/\sqrt N$, typically of the order $\sim$ mT when translated to an effective magnetic field.
Projecting this Hamiltonian to the qubit subspace yields
\begin{equation}
\hat{H}_\text{hf,q}
= \delta K^z \sin \frac{\theta}{2}\, \hat \sigma_x,
\label{eq:hfq}
\end{equation}
where $\delta K^z = \frac{1}{2}(K^z_L - K^z_R)$ is a quasistatic random field gradient.
For the singlet-triplet qubit this gradient can be used for initialization along the $\pm\hat x$-axis of the Bloch sphere \cite{Petta2005}, but in general its random nature presents a main source of qubit decoherence.
Protocols how to control or suppress the gradient $\delta K^z$ could lead to significant improvement of the qubit coherence time.

\subsection{Transport-induced nuclear spin pumping: Qualitative picture}
\label{sec:Intuitive_picture}

In Ref.~\cite{spin-blockade-quenching} it was shown how such a gradient can get suppressed naturally in the presence of strong spin-orbit interaction, when the double dot is embedded in a transport setup.
We will first review here the intuitive picture of the underlying mechanism, as outlined in Ref.~\cite{spin-blockade-quenching}, and then show how it also works in the absence of spin-orbit interaction.
In the next sections we will support this with an analytic investigation and numerical simulations of the coupled electron-nuclear spin dynamics.

\begin{figure}[tbp]
	\centering
	\includegraphics[scale=.73]{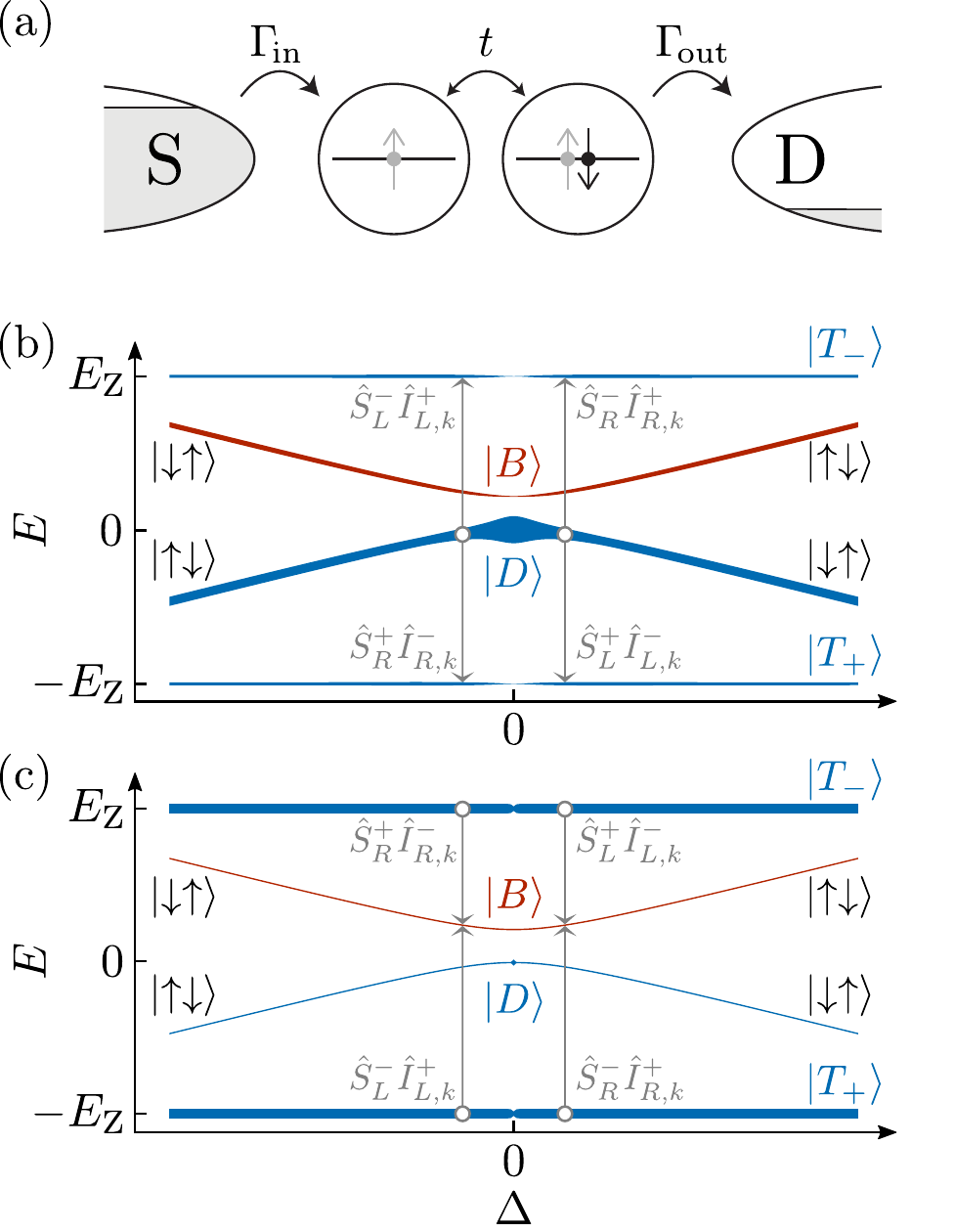}
	\caption{(a) The double quantum dot is tunnel coupled to source and drain reservoirs, and in the presence of a bias voltage electrons can flow from source to drain.
		Energy spectrum as a function of $\Delta$ for the (1,1) spin states: (b) with and (c) without spin-orbit coupling. The thickness of the lines indicates the occupation probability of the eigenstates as given by Eqs.~\eqref{eq:ppm}--\eqref{eq:p2}. Preferred electron-nuclear spin flip rates close to $\Delta =0$ are indicated by the gray arrows.}
	\label{fig:energy_spectrum}
\end{figure}

We assume the double dot to be connected in a linear arrangement to source and drain reservoirs, as sketched in Fig.~\ref{fig:energy_spectrum}a, and to be tuned close to the so-called ``triple point'' (where three stable charge regions meet) indicated by the red dot in Fig.~\ref{fig:model}a.
Then, a finite bias voltage over source and drain can give rise to a current through the system, via the transport cycle $(0,1)\to(1,1)\to(0,2)\to(0,1)$.
We assume that the system is tuned to the open regime, where the couplings to the reservoirs, characterized by the rates $\Gamma_{\rm in,out}$, are the largest relevant energy scales.
This ensures that the tunneling processes $(0,2) \to (0,1) \to (1,1)$ are effectively instantaneous, and the interesting dynamics happen during the transition $(1,1) \to (0,2)$ which involves the same five levels as before, $\{ \ket{T_{\pm,0}}, \ket{S}, \ket{S_{02}} \}$.

In the absence of spin-mixing processes, the only available transport path is $(0,1)\to \ket{S} \to \ket{S_{02}} \to(0,1)$ and population of one of the (1,1) triplet states results in spin blockade of the current.
The effect of SOI in this context is twofold:
(i) small inhomogeneities in the confining potential can result in different effective $g$-factors $g_{L,R}$ on the two dots,
and (ii) tunneling from one dot to the other can now be accompanied by a spin flip~\cite{Danon2009a}.
These two effects can be described by the Hamiltonian
\begin{align}
\hat H_{so} = {} & {} it^+\ket{T_-}\bra{S_{02}} - it^-\ket{T_+}\bra{S_{02}} 
\nonumber\\&
+ it_z\ket{T_0}\bra{S_{02}} + \Delta_{so} \ket{T_0}\bra{S} + {\rm H.c.},
\end{align}
where $t^\pm=\frac{1}{\sqrt{2}}\left(t_x\pm it_y\right)$, with the real vector ${\bf t}$ characterizing the spin-orbit induced spin-flip tunnel coupling, and $\Delta_{so} = \frac{1}{2}(g_L-g_R)\mu_{\rm B}B$ accounting for the difference in $g$-factors on the dots.
The magnitude of the vector ${\bf t}$ can be estimated as $\sim (d/l_{so}) t_s$, where $d$ is the distance between the two dots and $l_{so}$ the spin-orbit length in the direction of the interdot axis.

We see that SOI can lift the blockade of the polarized states $\ket{T_\pm}$.
But if the total Zeeman gradient $\Delta$ vanishes, $\Delta = \Delta_{so} + \delta K^z = 0$, the two unpolarized (1,1) states can still be combined into a bright state $\ket{B} = [t_s\ket{S} + it_z \ket{T_0}] / \sqrt{t_s^2 + t_z^2 }$ (that is coupled to $\ket{S_{02}}$ with strength $\sqrt{t_s^2 + t_z^2 }$) and a dark state $\ket{D} = [it_z\ket{S} + t_s \ket{T_0}] / \sqrt{t_s^2 + t_z^2 }$ (that is not coupled).
So in this case there is still one spin-blocked state left, $\ket{D}$, which, as a consequence, will be populated with high probability, whereas the other three states $\ket{T_\pm}$ and $\ket{B}$ have vanishing population.
Adding a finite Zeeman gradient $\Delta\neq 0$ mixes the states $\ket{T_0}$ and $\ket{S}$, and thus $\ket{B}$ and $\ket{D}$, lifting the blockade of $\ket{D}$ which results in a more evenly distributed population of the levels.
These observations are illustrated in Fig.~\ref{fig:energy_spectrum}b, where we show the energy spectrum of the four (1,1) states as a function of $\Delta $: The thickness of the lines indicates the relative occupation probabilities of the four states when embedded in a transport setup.
We have set $t_s = 0.6\,E_{\rm Z}$ and ${\bf t} = \{0.4,0.4,0.4 \} t_s$, and we assumed the escape rates of every state to be proportional to the modulo square of its total coupling to $\ket{S_{02}}$ given by $\hat H_t + \hat H_{so}$, which is valid in the limit of large $\Gamma_{\rm out}$.

Based on this, we can now develop a qualitative understanding of the resulting coupled electron-nuclear spin dynamics.
The hyperfine Hamiltonian (\ref{eq:hyperfine_hamiltonian}) contains terms $\hat{S}_d^\pm\hat{I}_{d,k}^\mp$ which can give rise to so-called spin flip-flop processes in which the electron on dot $d$ exchanges one unit of angular momentum with one of the nuclei in the dot, which changes the value of the effective nuclear field $K^z_d$ by a small amount.
A non-equilibrium electron spin polarization on the dots can thus be slowly transferred to the nuclear spin ensemble which, in turn, can influence the electron dynamics, potentially yielding an intricate feedback cycle.

To see if there is a preferred direction of nuclear spin polarization, we investigate the spin structure of the most strongly occupied electronic state:
At $\Delta = 0$ the state $\ket{D}$ contains equally large components of $\ket{\ua\da}$ and $\ket{\da\ua}$, i.e., $|\braket{{D}|{\ua\da}}|^2 = |\braket{{D}|{\da\ua}}|^2 = \frac{1}{2}$, where $\ket{\alpha\beta}$ denotes the (1,1) state with a spin-$\alpha$ electron on the left dot and a spin-$\beta$ electron on the right dot.
Due to these equal weights, all possible hyperfine-induced flip-flop processes are to first approximation equally likely, and the net nuclear spin flip rates on both dots thus vanish.
However, when $\Delta > 0$ the most strongly occupied state acquires a slightly $\da\ua$-polarized character (see Fig.~\ref{fig:energy_spectrum}b) and then the flip-flop processes caused by $\hat{S}_L^+\hat{I}_{L,k}^-$ and $\hat{S}_R^-\hat{I}_{R,k}^+$ (illustrated by the gray arrows in the figure) are more likely than the opposite ones.
This results in a net negative(positive) nuclear spin pumping rate in the left(right) dot, which reduces $\delta K^z$ and thus $\Delta $.
Similarly, we see that when $\Delta < 0$ the small polarization of the most strongly occupied state will drive $\delta K^z$ and thus $\Delta$ to larger values.
All together, this indeed suggests that the specific manifestation of spin blockade in the presence of strong SOI can result in a self-quenching of the Zeeman gradient over the dots.
The experimental results presented in Ref.~\cite{spin-blockade-quenching} were consistent with this picture.

Let us now turn to the limit of very weak SOI, where we set ${\bf t}= \Delta_{so} = 0$.
In that case we see that at $\Delta = 0$ there are three spin-blocked states, the (1,1) triplet states $\ket{T_{\pm,0}}$.
At this special point one thus finds an occupation probability of $\frac{1}{3}$ for each of the triplet states and zero for the coupled state $\ket{S}$.
But again, due to the symmetric polarization of all four states, there will be no net nuclear spin pumping at this point.
Away from the special point $\Delta = 0$, the Zeeman gradient mixes the states $\ket{S}$ and $\ket{T_0}$ and both unpolarized eigenstates end up having a finite coupling to $\ket{S_{02}}$, whereas the polarized triplets remain uncoupled.
This results in an occupation probability of approximately $\frac{1}{2}$ for $\ket{T_+}$ and $\ket{T_-}$ and zero for the two unpolarized states.
We first focus on the case $\Delta > 0$, where $\ket{D}$ evolves into a state with a slightly stronger $\da\ua$-component, whereas $\ket{B}$ acquires a slight $\ua\da$-character (see Fig.~\ref{fig:energy_spectrum}c).
Flip-flops from the blocked states can cause transitions to both unpolarized states, but due to its stronger coupling to $\ket{S_{02}}$ transitions to the state $\ket{B}$ at $\Delta = 0$ are favored.
This means that the flip-flop processes caused by $\hat{S}_L^+\hat{I}_{L,k}^-$ and $\hat{S}_R^-\hat{I}_{R,k}^+$ are most likely, which again result in a pumping of $\delta K^z$ toward smaller values of $\Delta$.
At $\Delta < 0$ a similar reasoning results in positive pumping of $\delta K^z$ toward higher values of $\Delta$.
So, we see that also in the case of vanishing SOI a naive qualitative investigation of the spin dynamics predicts a transport-induced self-quenching of the Zeeman gradient.

In the next two sections we will present analytic and numerical investigations that support the simple picture presented above.

\subsection{Analytic results}

\label{sec:Analytical_results}

We start by deriving evolution equations for the nuclear polarizations in the two dots, similar to those derived in Ref.~\cite{spin-blockade-quenching} but now including the effect of the strong couplings $\Gamma_{\rm in,out}$ in a more general way and not solely focusing on the case of strong SOI.
From the flip-flop rates we thus find, we derive an expression for the fluctuations around the stable point at $\Delta=0$ using a Fokker-Planck equation to describe the stochastic dynamics of the nuclear fields $K^z_{L,R}$.

We start from a time-evolution equation for the electronic density matrix (we use $\hbar=1$),
\begin{equation}
\frac{d\hat{\rho}}{dt}
= -i\big[\hat{H},\hat{\rho}\big]  + \bs{\Gamma}\hat{\rho},
\label{eq:Von_neumann}
\end{equation}
where $\hat H = \hat H_0 + \hat H_{so} + \delta K^z \big[ \ket{T_0}\bra{S} + \ket{S}\bra{T_0} \big]$.
We neglect all other components of ${\bf K}_{L,R}$ since they lead to small corrections that are of the order $K/E_{\rm Z}$, where $K$ is the typical magnitude of the nuclear fields.
The term $\bs{\Gamma}\hat\rho = - \frac{1}{2}\Gamma \{ \hat P_{02}, \hat\rho \} + \frac{1}{4} \Gamma (\mathbbm{1} - \hat P_{02} )\rho_{02,02}$ describes the transitions $\ket{S_{02}} \to (0,1) \to (1,1)$, using the projector onto the (0,2) singlet state $\hat P_{02} = \ket{S_{02}}\bra{S_{02}}$.

Assuming that the rate $\Gamma$ is the largest energy scale in (\ref{eq:Von_neumann}), we can separate the time scales of the part of $\hat \rho$ involving $\ket{S_{02}}$ and the part describing the dynamics in the (1,1) subspace.
This yields an effective Hamiltonian for that subspace
\begin{align}
\hat H^{(1,1)} = \left( \begin{array}{cccc}
E_{\rm Z} & 0 & 0 & 0 \\
0 & E_B & \Delta & 0 \\
0 & \Delta & 0 & 0\\
0 & 0 & 0 & -E_{\rm Z}
\end{array} \right),
\label{eq:h11}
\end{align}
written in the basis $\{ \ket{T_-}, \ket{B}, \ket{D}, \ket{T_+} \}$, where we assumed $g<0$ and $B>0$.
The projection onto the (1,1) subspace resulted in exchange terms of the form $(\hat H_{ex})_{ij} = 4\epsilon T_{ij} /(4\epsilon^2+\Gamma^2)$, with 
\begin{equation}
    T_{ij} = \bra{i}(\hat H_t+\hat H_{so})\ket{S_{02}}\bra{S_{02}}(\hat H_t+\hat H_{so})\ket{j},
\end{equation}
and thus $E_B=4\epsilon (t_s^2+t_{z}^2)/(4\epsilon^2+\Gamma^2)$.
Assuming that $E_{\rm Z}$ is much larger than all exchange corrections, we neglected the terms coupling $\ket{T_\pm}$ to $\ket{B,D}$.
The four (1,1) states also acquire a finite life time that can be characterized by the four decay rates $\Gamma_i = 4\Gamma T_{ii} /(4\epsilon^2+\Gamma^2)$, where we note that $\Gamma_+ = \Gamma_- \equiv \Gamma_t$.

Using (\ref{eq:h11}) and the decay rates $\Gamma_i$, we can write a time-evolution equation for $\hat\rho^{(1,1)}$ similar to (\ref{eq:Von_neumann}).
Solving $d\hat\rho^{(1,1)}/dt = 0$ we find the equilibrium density matrix, which can be written $\hat\rho^{(1,1)}_{\rm eq} = \sum_i p_i \ket{i}\bra{i}$ in the basis $\{ \ket{T_+},\ket{1},\ket{2},\ket{T_-}\}$, where
\begin{align}
\ket{1} = {} & {} \cos \frac{\theta}{2} \ket{D} + e^{i\varphi} \sin \frac{\theta}{2} \ket{B},\\
\ket{2} = {} & {}   \cos \frac{\theta}{2} \ket{B}  - e^{-i\varphi}\sin \frac{\theta}{2} \ket{D},
\end{align}
in terms of the angles $\varphi = {\rm arg}( -i\Gamma_B\Delta - 2E_{B} \Delta )$ and $\theta = {\rm arctan}\big(4|\Delta|/\sqrt{\Gamma_B^2 + 4E_{B}^2}\big)$.
The occupation probabilities $p_i$ of the four states read
\begin{align}
p_\pm = {} & {} \frac{4\Gamma_B \Delta^2}{\Gamma_t E_2^2 + 8 \Gamma_B\Delta^2 },\label{eq:ppm}\\
p_1 = {} & {} \frac{1}{2} - \frac{4\Gamma_B\Delta^2 - \frac{1}{2}\Gamma_t \sqrt{(4E_{B}^2+\Gamma_B^2)E_2^2}}{\Gamma_t E_2^2 + 8 \Gamma_B\Delta^2 }\label{eq:p1},\\
p_2 = {} & {} \frac{1}{2} - \frac{4\Gamma_B\Delta^2 + \frac{1}{2}\Gamma_t \sqrt{(4E_{B}^2+ \Gamma_B^2 )E_2^2}}{\Gamma_t E_2^2 + 8 \Gamma_B\Delta^2 }\label{eq:p2},
\end{align}
with $E_2 = \sqrt{4E_B^2 + \Gamma_B^2 + 16 \Delta^2}$.
In contrast to Ref.~\cite{spin-blockade-quenching}, we included the effect $\Gamma_\text{out}$ here, resulting in a different basis of unpolarized states $\ket{1,2}$. 

We now add the flip-flop terms in (\ref{eq:hyperfine_hamiltonian}) in a perturbative way where we use Fermi's golden rule to calculate the rates for the resulting nuclear spin flips.
Assuming for simplicity nuclear spin $\frac{1}{2}$~\footnote{Using a different nuclear spin, such as $\frac{3}{2}$ (as it is for both Ga and As), yields unimportant overall numerical prefactors of order 1.}, we write for the flip rates up and down on dot $d$
\begin{align}
\gamma^\pm_d = \frac{A^2}{4 N^2} N_d^\mp \sum_{i,j} p_i\frac{\Gamma_j}{E^2_{\rm Z}}
| \bra{j} \hat S^\mp_d \ket{i} |^2 + \gamma N_d^\mp,\label{eq:flip}
\end{align}
where $N_d^\pm$ is the number of nuclei with spin $\pm \frac{1}{2}$ on the dot.
The factor $\Gamma_j / E_{\rm Z}^2$ accounts for the finite life time of the final electronic state $\ket{j}$, assuming a Lorentzian level broadening in the limit $E_{\rm Z} \gg \Gamma_j$.
We also added a term that describes random nuclear spin flips with a rate $\gamma$ to account phenomenologically for the slow relaxation of the nuclear spins to their fully-mixed equilibrium state.

We can translate these flip rates to evolution equations for the dot polarizations $P_d = (N^+_d - N^-_d) / N$.
For the polarization gradient $P_\Delta = \frac{1}{2}\left(P_L-P_R\right)$ and the average polarization $P_\Sigma = \frac{1}{2}\left(P_L+P_R\right)$ we find
\begin{align}
\frac{d P_\Delta}{dt} \! = &
- \!\bigg[ F(\Delta) +\frac{1}{\tau} \bigg]P_\Delta \! - \! \frac{2F(\Delta)E_B\Delta }{E_B^2 + \frac{1}{4}\Gamma_B^2 + 4 \Delta^2},
\label{eq:analytic_gradient_result_DQD}\\
\frac{dP_\Sigma}{dt} \!
= &  - \! \bigg[ F(\Delta) +\frac{1}{\tau} \bigg]P_\Sigma,
\label{eq:analytic_average_result_DQD}
\end{align}
with
\begin{equation*}
F(\Delta) = \frac{A^2}{4N^2E_{\rm Z}^2}\frac{\Gamma_t^2(4E_B^2 + \Gamma_B^2 + 16 \Delta^2) +4\Gamma_B^2\Delta^2}{\Gamma_t (4E_B^2 + \Gamma_B^2 + 16 \Delta^2) + 8\Gamma_B\Delta^2},
\end{equation*}
and $1/\tau = 2\gamma/N$ the phenomenological relaxation rate of the polarizations, usually $\tau \sim$ 1--10~s.
We note that these equations are non-linear, since $\Delta = \Delta_{so} + \delta K^z = \Delta_{so} + (A/2)P_\Delta$.

\begin{figure}[tbp]
	\centering
	\includegraphics[scale=.73]{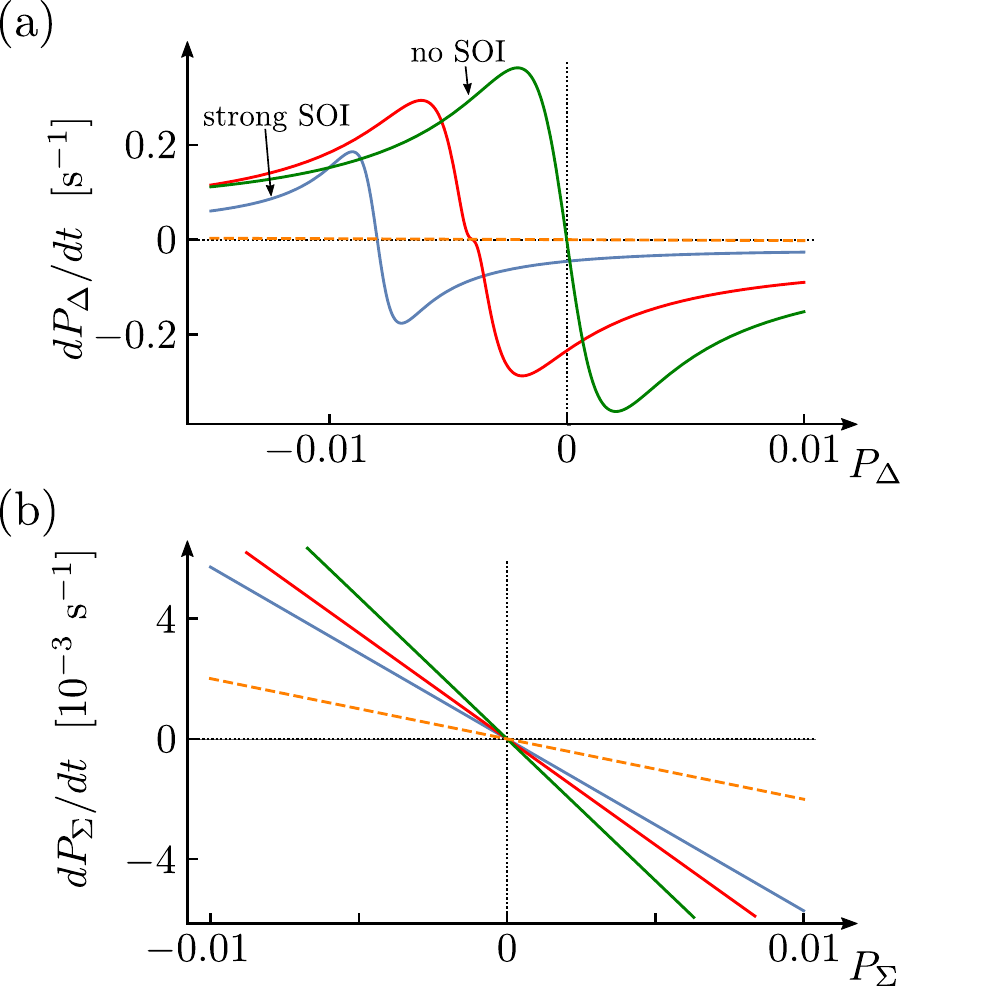}
	\caption{
		Pumping curves for the polarization gradient and average polarization, as given by Eqs.~(\ref{eq:analytic_gradient_result_DQD}) and (\ref{eq:analytic_average_result_DQD}).
		(a) $dP_\Delta/dt$ as a function of $P_\Delta$.
		(b) $dP_\Sigma/dt$ as a function of $P_\Sigma$.
		In both plots we show three curves: without SOI (green), with intermediate SOI (red), and with strong SOI (blue); see the main text for the parameters used.
		As reference, we also added the result without any spin pumping, i.e., with $F(\Delta)=0$ (orange dashed line).
	}
	\label{fig:dpdt}
\end{figure}

From Eqs.~(\ref{eq:analytic_gradient_result_DQD}) and (\ref{eq:analytic_average_result_DQD}) we see that both polarizations acquire effectively an enhanced relaxation rate, $\tau^{-1} \to \tau^{-1} + F(\Delta)$, which does depend on $P_\Delta$ but always drives the polarizations toward zero.
Furthermore, \eqref{eq:analytic_gradient_result_DQD} has an extra term that pumps the polarization gradient to the point where the \emph{total} Zeeman gradient $\Delta$ is zero.
For typical parameters, where $E_B \sim \Gamma_B \ll A$, this term dominates and the result is a stable polarization close to $\Delta = 0$.
In the limit of vanishing SOI, we can set $\Gamma_t \to 0$ and then find $F(\Delta) = A^2 \Gamma_B / 8 N^2 E_{\rm Z}^2$.
These results are illustrated in Fig.~\ref{fig:dpdt}, where we plot (a) $dP_\Delta/dt$ as a function of $P_\Delta$ and (b) $dP_\Sigma/dt$ as a function of $P_\Sigma$ for three different strengths of SOI (green, red, and blue lines) as well as without any spin pumping (orange dashed line).
We used $A = 250~\mu$eV, $E_{\rm Z} = 5~\mu$eV, $N = 4 \times 10^5$, and $\tau = 5$~s.
For the curve without SOI (green) we used $E_B = 0.5~\mu$eV, $\Gamma_B = 0.25~\mu$eV, and $\Gamma_t = \Delta_{so} = 0$.
The other two curves have $\Gamma_t = 0.01~\mu$eV, $\Delta_{so} = 0.5~\mu$eV (red) and $\Gamma_t = 0.0625~\mu$eV, $\Delta_{so} = 1~\mu$eV (blue).
In these two cases, we adjusted $\Gamma_B$ and $E_B$ such that the total coupling $\sqrt{t_s^2 + |{\bf t}|^2 }$ remains constant; this amounts to assuming that the SOI ``converts'' part of the tunnel coupling to a non-spin-conserving coupling but it does not affect the total coupling energy.
In the next section, we will show that these analytic results also agree well with numerical simulations of the dynamics of the polarizations, see Fig.~\ref{fig:numerical_simulation}a.

\begin{figure}[tbp]
	\centering
	\includegraphics[scale=.65]{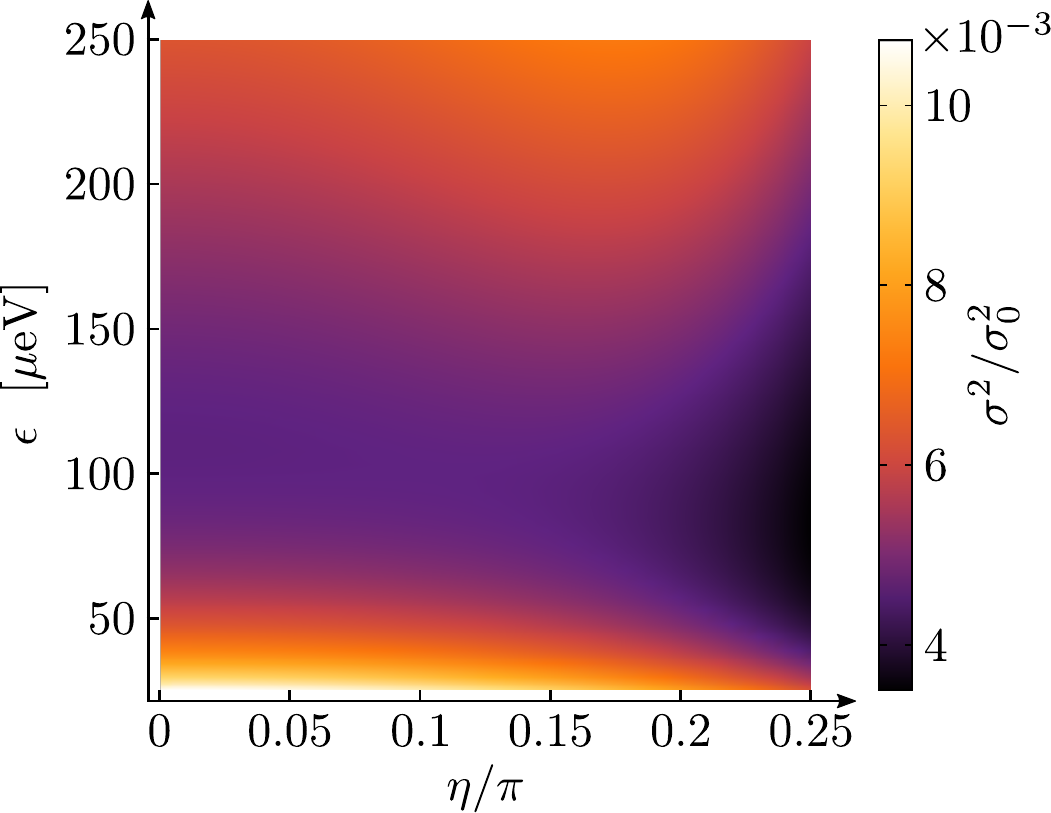}
	\caption{Suppression of the fluctuations of the nuclear field gradient, as given in (\ref{eq:fluctuations}), as a function of $\epsilon$ and $\eta$, where $\eta$ characterizes the strength of the SOI.
		See the main text for the parameters used and the exact definition of $\eta$.
	}
	\label{fig:sigma}
\end{figure}

Finally, we investigate the stochastic fluctuations of the polarization gradient around the stable point using a Fokker-Planck equation to describe the (time-dependent) probability distribution function ${\cal P}(n,t)$, where the integer $n = N P_\Delta$ labels the allowed polarization gradients~\cite{Danon2009,Vink2009}.
Going to the continuum limit, we can find the equilibrium distribution function to be
\begin{equation}
{\cal P}(P_\Delta) = \exp \left\{ \int^{P_\Delta} dP_\Delta'\, 2N \frac{\gamma_\Delta^+ - \gamma_\Delta^-}{\gamma_\Delta^+ + \gamma_\Delta^-} \right\},
\end{equation}
where $\gamma^\pm_\Delta = \frac{1}{2}(\gamma_L^\pm - \gamma_R^\pm)$ in terms of the flip rates as written in (\ref{eq:flip}).
The slope of the integrand close to the points where $\gamma_\Delta^+ - \gamma_\Delta^-=0$ can thus be used to estimate the equilibrium r.m.s.~deviation of $P_\Delta$ from those stable points.
In the absence of pumping, i.e., for $F(\Delta)\to 0$, we find a peak in the distribution around the point $P_\Delta = 0$ with a variance $\sigma_0^2 = 1/2N$.
Including pumping, and assuming that the second term in (\ref{eq:analytic_gradient_result_DQD}) dominates around the stable point, we find a peak in ${\cal P}(P_\Delta)$ at $P_\Delta \approx -2 \Delta_{so} / A$, where
\begin{equation}
\sigma^2 \approx
\sigma^2_0
\frac{E_B^2 + \frac{1}{4}\Gamma_B^2}{A E_B}\left(1 + 8 \frac{E_{\rm Z}^2N^2}{A^2\tau}\frac{\Gamma_B + 2\Gamma_t}{\Gamma_B^2 + 4\Gamma_t^2}\right).
\label{eq:fluctuations}
\end{equation}
In Fig.~\ref{fig:sigma} we show the resulting suppression of the fluctuations $\sigma^2/\sigma_0^2$ as a function of detuning $\epsilon$ and strength of the SOI, parameterized by $\eta$, where we fixed the total tunnel coupling to $t = 7.5~\mu$eV and then used $t_x^2 + t_y^2 = t^2 \sin^2 \eta$ and $t_z^2 + t_s^2 = t^2 \cos^2 \eta$.
In this way, $\eta = 0$ corresponds to having no SOI and $\eta \sim \pi/4$ to strong SOI.
We further used $A = 250~\mu$eV, $E_{\rm Z} = 12.5~\mu$eV, $\Gamma = 75~\mu$eV, $N = 4 \times 10^5$, and $\tau = 5$~s.
For these parameters we observe a significant suppression of the fluctuations in the whole range we plotted.
We see that the suppression is most effective for strong SOI (where $\eta \to \pi/4$), but still of the same order of magnitude in the absence of SOI (where $\eta = 0$).

\subsection{Numerical simulations}
\label{sec:Numerical_simulation}

We complement our analytic results with a numerical simulation of the electron-nuclear spin dynamics, discretizing time in small steps of $\Delta t$.
We start with two initial polarizations $P_L(0)$ and $P_R(0)$ on the two dots and then solve for the eigenvalues $\varepsilon_i$ and eigenmodes $\hat \rho_i$ of the superoperator $\Lambda$ that describes the coherent evolution and decay of the density matrix,
\begin{align}
\Lambda \hat \rho = -i\big[\hat{H},\hat{\rho}\big] - \frac{1}{2} \big\{ \hat \Gamma, \hat \rho \big\},
\end{align}
where $\hat \Gamma$ is a diagonal matrix containing the decay rates of the five basis states~\footnote{We added an infinitesimal decay rate of $10^{-9}~\mu$eV to all (1,1) states to avoid singularities.}.
Each of the 25 eigenmodes of $\Lambda$ can then be written as $\hat \rho_i = \ket n \bra m$ where $\ket n$ and $\ket m$ are picked from a (new) five-dimensional basis.
The corresponding eigenvalue $\varepsilon_i$ has the form $\varepsilon_i = -i(E_n - E_m) - \frac{1}{2}(\gamma_n + \gamma_m)$ where $E_{n,m}$ and $\gamma_{n,m}$ give the effective energies and decay rates of the two states $\ket n$ and $\ket m$.
From knowing all $\varepsilon_i$ and $\hat \rho_i$ we can thus derive the appropriate basis states, their effective energies, and their decay rates.
To find the steady-state occupation probabilities for these five states, we evaluate their weight in the (1,1) subspace, $w_n = \bra{n} (\mathbbm{1} - \hat P_{02} ) \ket{n}$, from which the occupation probabilities follow as $p_n = w_n\gamma_n^{-1} / \sum_i w_i \gamma_i^{-1}$.

Now we have all ingredients we need to evaluate the spin flip rates on both dots.
We rewrite Eq.~(\ref{eq:flip}) including the detailed dependence on all energy differences and decay rates,
\begin{align}
\gamma^\pm_d = {} & {} \frac{A^2}{N^2}
\sum_{i,j} \frac{p_i(\gamma_i + \gamma_j)| \bra{j} \hat S^\mp_d \ket{i} |^2}{4(E_i - E_j)^2 + (\gamma_i + \gamma_j)^2}  N_d^\mp + \gamma N_d^\mp.
\label{eq:flip_full}
\end{align}
Then we pick random numbers of spin-flip events $k_d^\pm$ on both dots and in both directions, using a Poisson distribution $(\gamma_d^\pm \Delta t)^{k_d^\pm} e^{\gamma_d^\pm \Delta t} / (k_d^\pm)!$, and we update the polarizations $P_d(\Delta t) = P_d(0) + (2/N) (k_d^+ - k_d^-)$.
This process can then be repeated as many times as desired to simulate the evolution of $P_{L,R}(t)$ over longer times.
We note that we make sure that $\Delta t$ is small enough so that most of the $k_d^\pm$ turn out 0 or 1.

\begin{figure}[tb]
	\centering
	\includegraphics[width=0.48\textwidth]{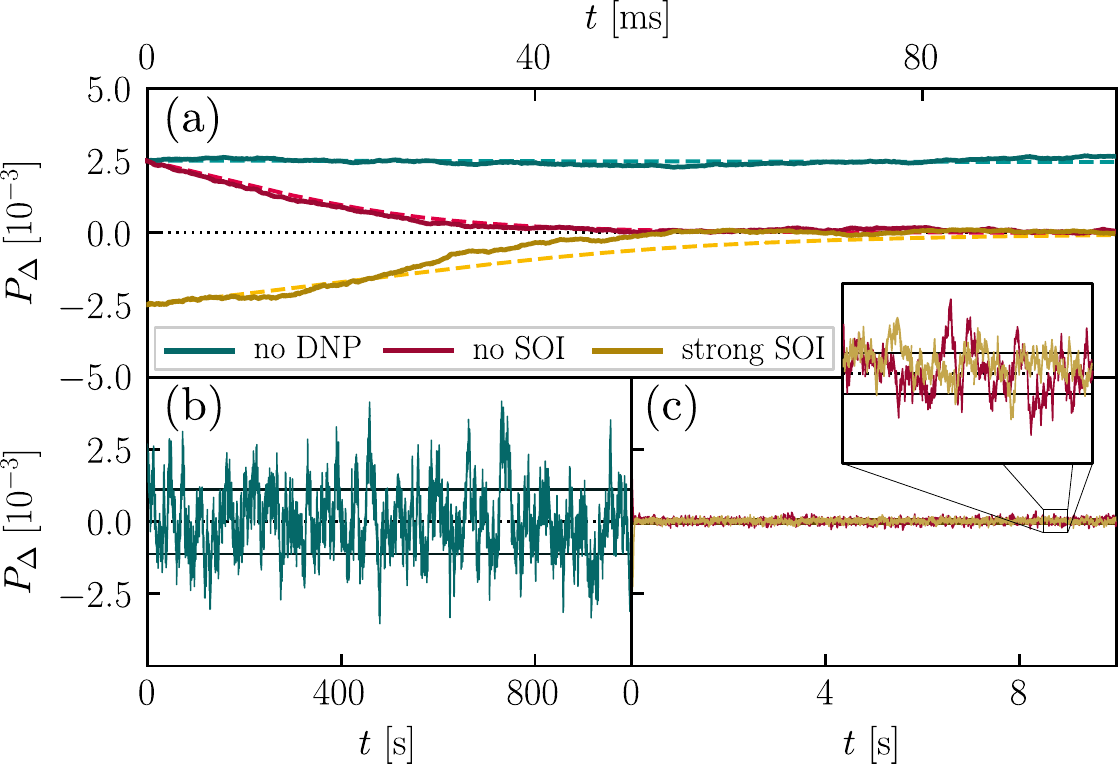}
	\caption{Simulation of the polarization gradient $P_\Delta(t)$ without SOI (red solid lines), strong SOI (yellow solid lines), and without hyperfine-induced spin pumping ($A = 0$, green solid lines).
		The dashed lines show the solution of Eq.~(\ref{eq:analytic_gradient_result_DQD}) using the same parameters.
		(a) Short-time evolution. Note that we used different initial conditions for clarity: $P_\Delta(0) = 0.0025$ for the red and green lines and $P_\Delta(0) = -0.0025$ for the yellow line; we always set $P_\Sigma(0)=0$.
		(b) Long-time evolution for $A=0$. The horizontal black lines indicate $\pm \sigma_0$.
		(c) Long-time evolution in the presence of spin pumping. The horizontal black lines now show $\pm \sigma$ as found from Eq.~(\ref{eq:fluctuations}) (see inset).
		See the main text for all other parameters used.}
	\label{fig:numerical_simulation}
\end{figure}

We show the results of our simulations as solid lines in Fig.~\ref{fig:numerical_simulation}, where we plot $P_\Delta(t)$ for three different cases: (i) strong SOI, where $t_{x,y,z} = 3.12~\mu$eV and $t = 5.21~\mu$eV (yellow), (ii)~no SOI, with $t_{x,y,z} = 0$ and $t = 7.5~\mu$eV (red), and (iii) no hyperfine interaction (green).
The other parameters used were $A=125~\mu$eV, $E_{\rm Z}=12.5~\mu$eV, $\delta=100~\mu$eV, $\Gamma=75~\mu$eV, $N=4\times10^5$, $\tau=5$~s, and $\Delta t = 10~\mu$s.
We used as initial conditions $P_\Delta(0) = 0.0025$ (red and green), $P_\Delta(0) = -0.0025$ (yellow), and $P_\Sigma(0) = 0$ (always).
We note that, in order to make comparison more straightforward, we set $\Delta_{so} = 0$ in all cases, including the case of strong SOI.

In Fig.~\ref{fig:numerical_simulation}(a) we show the first 0.1~s of the evolution.
We see that the hyperfine interaction accelerates the dynamics of the polarizations and tends to suppress the gradient to zero.
We added dashed lines that show time-dependent solutions of Eq.~(\ref{eq:analytic_gradient_result_DQD}), which indeed seems to predict the average dynamics of the polarization gradient to reasonable accuracy.
In Figs.~\ref{fig:numerical_simulation}(b,c) we show longer time traces to illustrate the magnitude of the fluctuations around the stable point $P_\Delta = 0$.
In Fig.~\ref{fig:numerical_simulation}(b) the fluctuations are clearly much larger than in \ref{fig:numerical_simulation}(c), which is what we expected.
The horizontal lines show the magnitude of the fluctuations as predicted by Eq.~(\ref{eq:fluctuations}): For the parameters used we find $\sigma_0 = 1.1 \times 10^{-3}$ (to be compared with the green trace), and $\sigma = 7.8\times 10^{-5}$ (red trace) and $\sigma = 7.5\times 10^{-5}$ (yellow trace).

In both simulations that include spin pumping (red and yellow lines) the average polarization $P_\Sigma$ tends to drift to negative values, stabilizing at $\sim -0.02$.
This can be understood in qualitative terms from Fig.~\ref{fig:energy_spectrum}(b,c):
With strong SOI [Fig.~\ref{fig:energy_spectrum}(b)] the state $\ket{T_+}$ decays more efficiently than $\ket{T_-}$ since it is closer in energy to $\ket{S_{02}}$ and $\Gamma$ is finite.
This makes in general spin flips from $\ket D$ slightly more likely to happen to $\ket{T_+}$, resulting in a net average transfer of negative angular momentum to the nuclear spins.
Without SOI [Fig.~\ref{fig:energy_spectrum}(c)], the bright state $\ket B$ is closer in energy to $\ket{T_-}$ than to $\ket{T_+}$ (assuming $\delta >0$), resulting in the flip rate $\ket{T_-} \to \ket B$ to be larger than $\ket{T_+}\to \ket B$.
This should indeed also result in a small net negative pumping of the average polarization.
These effects are not reflected in Eq.~(\ref{eq:analytic_average_result_DQD}) since in that Section we neglected all energy differences in the (1,1) subspace compared to $E_{\rm Z}$, which, in turn, was assumed negligible compared to $\Gamma$.

\subsection{Conclusion}

We found that embedding a double quantum dot in the spin-blockade regime in a transport setup, the flow of electrons induces dynamic nuclear spin polarization that tends to suppress the polarization gradient over the two dots.
This mechanism not only works in the case of strong SOI, but also with weak SOI or in the absence of SOI.
We derived simple analytic equations to describe the dynamics of the polarization gradient (which we corroborated with numerical simulations), and we found that, over a large range of parameters, the r.m.s.\ value of the random polarization gradient can be suppressed by one to two orders of magnitude.
This could present a straightforward way to extend the coherence time of double-dot-based spin qubits.

\section{Exchange-only qubit}\label{sec:triple_dot}
\subsection{The qubit}\label{sec:XO_qubit}
Exchange-only qubits are usually hosted in a linear triple quantum dot, with one electron in each dot.
The eight-dimensional (1,1,1) subspace consists of one spin quadruplet $\ket{Q}$ and two doublets $\ket{D_1}$ and $\ket{D_2}$.
An external magnetic field lifts the degeneracy of states with different total $S_z$, and when the system is then tuned close to the border of the (1,1,1) region, exchange effects due to finite interdot tunneling can lift the remaining degeneracies.
The qubit is then commonly defined in the two doublet states with spin projection $S_z = +\frac{1}{2}$, and turns out to be fully controllable via electric fields only.

\begin{figure}[tb]
	\centering
	\includegraphics[width=0.48\textwidth]{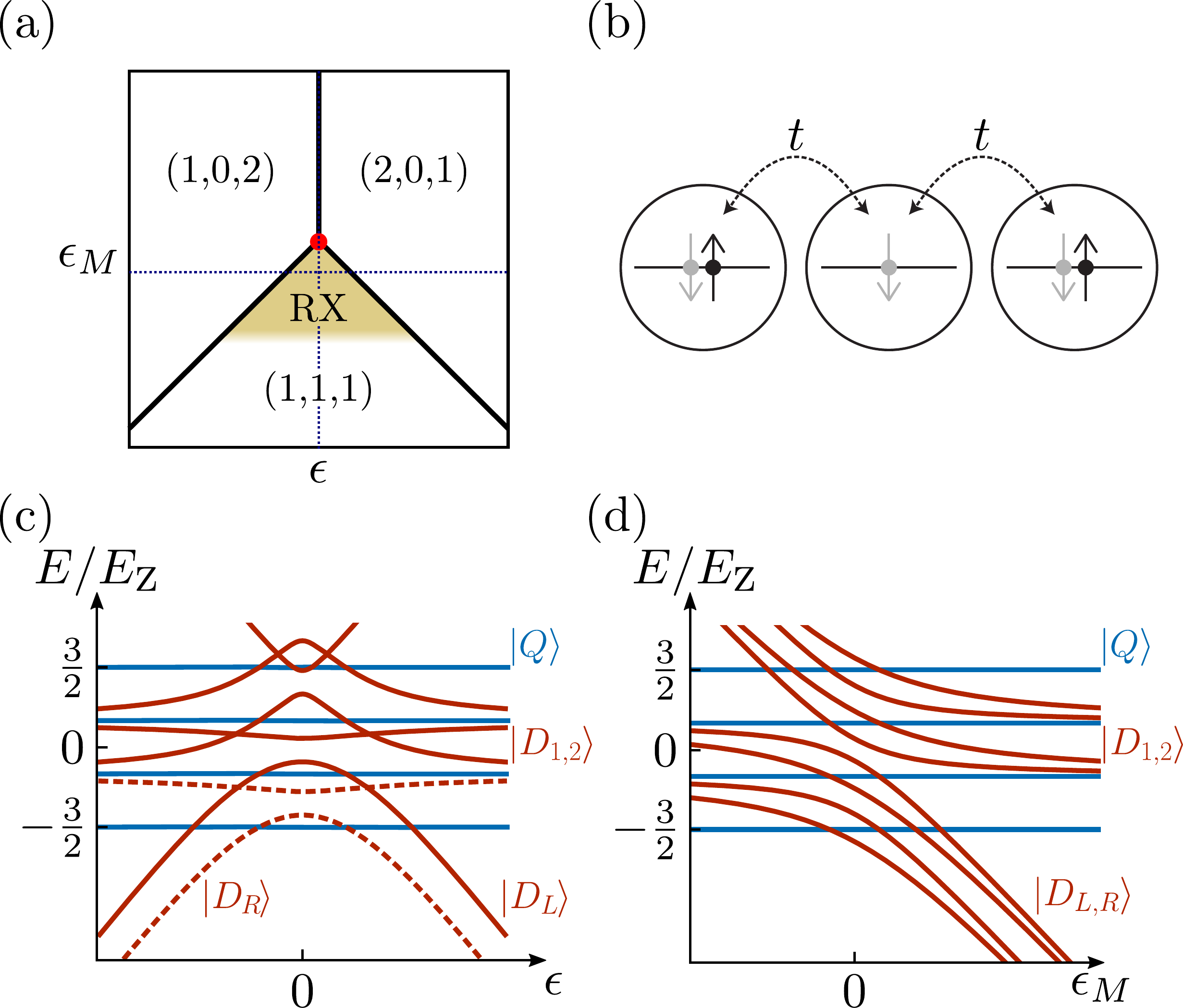}
	\caption{(a) Sketch of the charge stability diagram of a linear triple quantum dot tuned close to the (1,1,1)--(1,0,2)--(2,0,1) triple point. (b) Cartoon of the setup. (c,d) Lowest part of the spectrum along the horizontal and vertical dashed line in (a), respectively.}
	\label{fig:model_3dot}
\end{figure}

In Fig.~\ref{fig:model_3dot}(a) we sketch the charge stability diagram close to the (1,1,1)--(1,0,2)--(2,0,1) triple point, as a function of the two tuning parameters $\epsilon=\frac{1}{2}(V_R-V_L)$ and $\epsilon_M= V_C-\frac{1}{2}(V_L+V_R)$, where $V_{L,C,R}$ denote the gate-induced potentials on the left, central, and right dot, respectively.
We include energy offsets such that the triple point is defined to be at $(\epsilon_M,\epsilon) = (0,0)$.
In this regime, the low-energy part of the spectrum consists of 12 states:
In addition to the eight (1,1,1) states mentioned above, we also need to include a doublet $\ket{D_L}$ in a (2,0,1) configuration and a doublet $\ket{D_R}$ in a (1,0,2) configuration.

We can then write a similar Hamiltonian as before,
\begin{equation}
\hat{H}_0 = \hat{H}_e + \hat{H}_t + \hat{H}_{\rm Z}.
\label{eq:tqd_Hamiltonian}
\end{equation}
Now we have
\begin{align}
\hat H_e = \sum_{\alpha = \pm} \Big\{
{} & {} -(\epsilon_M + \epsilon) \ket{D_L^\alpha}\bra{D_L^\alpha}
\nonumber\\ {} & {} 
\ -(\epsilon_M - \epsilon) \ket{D_R^\alpha}\bra{D_R^\alpha} \Big\},
\end{align}
where $\alpha = \pm$ labels the spin projection $S_z = \pm \frac{1}{2}$ of the doublet state.
The tunneling Hamiltonian is
\begin{align}
\hat H_t = \frac{t}{2} \sum_{\alpha = \pm} \alpha \Big\{ {} & {}  \sqrt 3 \ket{D_1^\alpha} \big[ \bra{D_R^\alpha} - \bra{D_L^\alpha} \big] \nonumber\\{} & {}
+\ket{D_2^\alpha}\big[ \bra{D_R^\alpha} + \bra{D_L^\alpha} \big] \Big\} + \text{H.c.}
\end{align}
where we assumed the left and right tunneling couplings equal, for simplicity.
The Zeeman term is
\begin{align}
\hat H_{\rm Z} = g\mu_{\rm B} B \hat S_z^{\rm tot},
\end{align}
in terms of the total spin-$z$ projection operator for the three electrons.

In the region marked `RX' in Fig.~\ref{fig:model_3dot}(a) the central electron can become delocalized over the three dots [see Fig.~\ref{fig:model_3dot}(b)], yielding relatively strong exchange effects.
To illustrate, we sketch in Fig.~\ref{fig:model_3dot}(c) the spectrum of $\hat H_0$ along the dotted line in (a), where we set $t = 3\, E_{\rm Z}$.
The two dashed lines (the lowest doublet states with $S_z^{\rm tot} = +\frac{1}{2}$) form the qubit subspace, where  $\ket{1}=\ket{D_{2}^+}$ and $\ket{0}=\ket{D_{1}^+}$ at $\epsilon = 0$.
Close to that point, the projected qubit Hamiltonian is
\begin{equation}
    \hat{H}_q = \frac{J}{2}\hat{\sigma}_z - \frac{\sqrt 3 j}{2}\hat{\sigma}_x,
\end{equation}
with $J = \frac{1}{2}(J_L+J_R)$ and $j=\frac{1}{2}(J_L-J_R)$, in terms of the exchange energies $J_{L,R}$ associated with virtual tunneling to the left or right dot, respectively.
To lowest order in $t$ [valid not too close to the borders of the (1,1,1) region] we have $J_{L,R} = -t^2/(\epsilon_M\pm\epsilon)$.
From this it is clear that the exchange-only qubit allows for electric control of rotations around two different axes of the Bloch sphere, by tuning $J$ and $j$ through $\epsilon$ and $\epsilon_M$, whereas the singlet-triplet qubit offered electric control over only one axis.

As in the double-dot system, the main effect of the hyperfine interaction with the nuclear spin bath can be described on a mean-field level using three random effective nuclear fields,
\begin{equation}
    \hat{H}_\text{hf,mf}
    = \mathbf{K}_L\cdot\hat{\mathbf{S}}_L + \mathbf{K}_C\cdot\hat{\mathbf{S}}_C + \mathbf{K}_R\cdot\hat{\mathbf{S}}_R.
\end{equation}
Projected onto the qubit subspace, this yields
\begin{align}
\hat H_{\rm hf,q} = -\frac{2}{3} \delta K_M^z\hat \sigma_z - \frac{1}{\sqrt 3} \delta K_{LR}^z \hat \sigma_x,
\end{align}
where $\delta K^z_M=-\frac{1}{2}(\delta K^z_{LC} - \delta K^z_{CR})$ and $\delta K^z_{LR}=\frac{1}{2}\left(K_L^z-K_R^z\right)$, in terms of the field gradients $\delta K_{ij}^z = \frac{1}{2}(K^z_i - K^z_j)$ over neighboring dots.
We thus see that, also in this case, the random nuclear fields can be an important source of qubit decoherence.
Besides, the quadruplet state $\ket{Q_{+1/2}}$ that cannot be split off by increasing the external field $B$ is coupled to the states $\ket 0$ and $\ket 1$ through the same gradients $\delta K^z_M$ and $\delta K^z_{LR}$, which can thus cause leakage out of the qubit subspace.
To be able to control or suppress the field gradients could therefore again dramatically increase the qubit quality.

\subsection{Transport-induced nuclear spin pumping: Qualitative picture}

Inspired by our findings for the double dot, we now investigate possibilities to suppress the nuclear field gradients by running a current through the system while tuning it to some sort of spin-blockade regime.
In contrast to the double dot setup, there are several different types of spin blockade in a linear triple dot~\cite{Hsieh2012a}, which differ in the geometry of drains and sources and relative detuning of the three dots.
In a simplest setup where source and drain are attached to the outer dots, all regimes of spin blockade effectively behave as a double dot connected to one isolated dot containing one ``inert'' spin.
Transport through such a setup would thus only suppress the field gradient between the two interacting dots.

\begin{figure}[tbp]
	\centering
	\includegraphics[width=0.45\textwidth]{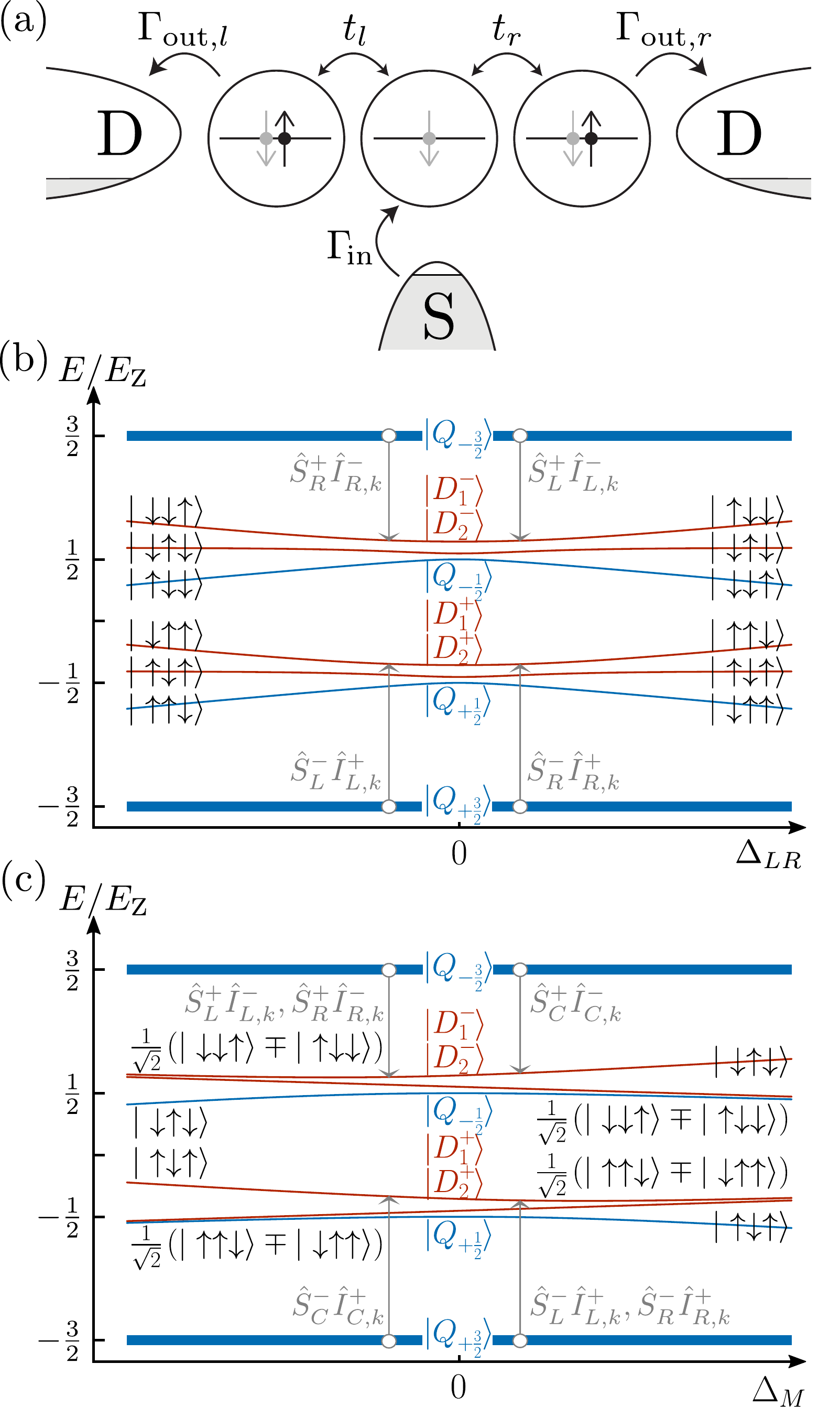}
	\caption{(a) The central dot is connected to a source and the two outer dots are connected to drains; an applied bias voltage then enables electrons to flow through the system to either of the drains.
	(b,c) Spectrum of the (1,1,1) states in the absence of SOI, as a function of the gradients $\Delta_{LR}$ (b) and $\Delta_M$ (c), where the thickness of the lines indicates the occupation probabilities. Preferred spin-flip rates are indicated by gray arrows.}
	\label{fig:energy_spectrum_3dot}
\end{figure}

To address both field gradients we use a setup where the source is connected to the \emph{central} dot and both of the outer dots are connected to a drain, see Fig.~\ref{fig:energy_spectrum_3dot}(a).
Applying a source--drain bias voltage in vicinity of the triple point shown in Fig.~\ref{fig:model_3dot}(a) can then give rise to a current through the system via the two transport cycles $(1,1,1)\to (2,0,1)/(1,0,2)\to (1,0,1)\rightarrow(1,1,1)$.
Again we will assume that the system is in the open regime where the rates $\Gamma_\text{in,out}$ are the largest energy scales, such that the interesting dynamics happen during the $(1,1,1) \to (2,0,1)/(1,0,2)$ transitions, which involves the 12 spin states discussed above.
For simplicity, we will assume a symmetric situation, where $\epsilon=0$ and $\epsilon_M > 0$ [see Fig.~\ref{fig:model_3dot}(d)], $t_l = t_r$, and $\Gamma_{{\rm out},l} = \Gamma_{{\rm out},r}$.

In absence of spin-mixing processes, the only (1,1,1) states that couple to $\ket{D_L}$ and $\ket{D_R}$ are the doublets $\ket{D_{1,2}}$, and the current is spin blocked in either of the four quadruplet states.
This blockade may be lifted by SOI, which affects the system in the same way as before:
(i) variations in the effective $g$-factor over the dots yield spin-orbit-induced Zeeman gradients $\Delta_{so,ij} = \frac{1}{2}(g_i-g_j)\mu_{\rm B}B$ and (ii) tunneling between dots can be accompanied by a spin flip.
It is easy to show that, in contrast to the double-dot case, in the presence of SOI there are no dark states, even when all total Zeeman gradients $\Delta_{ij} = \Delta_{so,ij}+\delta K^z_{ij}$ are zero.
SOI thus always fully lifts the spin blockade and competes with the flip-flop terms in the hyperfine interaction, thereby reducing the efficiency of spin pumping.
We will below only focus on the case without SOI, which is experimentally also most relevant since with strong SOI there is no spin blockade that can be used for initialization or read-out.

Let us now develop an intuitive picture of the electron-nuclear spin dynamics in this spin-blockade situation, similar to the discussion in Sec.~\ref{sec:Intuitive_picture}.
When the gradients $\Delta_{LR}$ and $\Delta_M$ are zero, the electrons are trapped in one of the four quadruplet states with equal probability $\frac{1}{4}$.
As before, due to the symmetric spin structure of all states at this point there will be no net spin pumping.
A non-zero gradient mixes states with the same total $S_z^{\rm tot}$, giving all six states with $S_z^{\rm tot} = \pm \frac{1}{2}$ a finite coupling to $\ket{D^\pm_{L,R}}$, whereas the two fully polarized quadruplets remain spin blocked, each with occupation probability $\frac{1}{2}$.
For small gradients, the doublets have a much larger coupling to $\ket{D^\pm_{L,R}}$ than $\ket{Q_{\pm 1/2}}$ and spin-flip processes are thus dominated by transitions from $\ket{Q_{\pm 3/2}}$ to a doublet state.

We first show that transitions to $\ket{D^\pm_2}$ do not contribute strongly to spin pumping.
When $\Delta_{LR}\neq0$, the states $\ket{D_{2}^\pm}$ develop a dominating $\ua\da\ua$- and $\da\ua\da$-character, respectively, see Fig.~\ref{fig:energy_spectrum_3dot}(b).
This results in an increased spin-flip rate $\gamma_C^+$ from transitions $\ket{Q_{+3/2}} \to \ket{D_2^+}$ as well as an increased rate $\gamma_C^-$ from $\ket{Q_{-3/2}} \to \ket{D_2^-}$.
One thus does not expect a strong net effect.
For $\Delta_M\neq0$ the situation is similar: $\ket{D_2^+}$($\ket{D_2^-}$) gains a larger weight of $\ua\ua\da$ and $\da\ua\ua$ ($\da\da\ua$ and $\ua\da\da$).
The spin-flip rates from $\ket{Q_{+3/2}} \to \ket{D_2^+}$ and $\ket{Q_{-3/2}} \to \ket{D_2^-}$ are thus affected in a symmetric way and there is no net spin pumping.

The doublet states $\ket{D_1^\pm}$, however, have the largest coupling to the outgoing states $\ket{D^\pm_{L,R}}$, and effectively pump the field gradients toward zero. 
For a positive gradient $\Delta_{LR}>0$, the state $\ket{D_{1}^+}$($\ket{D_{1}^-}$) evolves into a state with slight $\ua\ua\da$($\ua\da\da$)-character, see Fig.~\ref{fig:energy_spectrum_3dot}(b).
This increases $\gamma_R^+$($\gamma_L^-$) and thus drives $\Delta_{LR}$ toward lower values.
For a negative gradient $\Delta_{LR} < 0$, the situation is exactly opposite, again driving $\Delta_{LR}$ to zero.
A similar argument holds for the other gradient $\Delta_M$:
When $\Delta_M>0$, the state $\ket{D_1^-}$ gets a slight $\da\ua\da$-character and $\ket{D_1^+}$ obtains stronger $\da\da\ua$- and $\ua\da\da$-components, see Fig.~\ref{fig:energy_spectrum_3dot}(c).
This increases the rates $\gamma_L^+$, $\gamma_C^-$, and $\gamma_R^+$, thereby effectively reducing $\Delta_M$.
For $\Delta_M<0$ the situation is again opposite, yielding a positive pumping of $\Delta_M$.

\subsection{Analytic results}
We now use the same approach as in Sec.~\ref{sec:Analytical_results} to derive time-evolution equations for the three nuclear polarizations, valid for small $P_d$.
The time-evolution equation for the electronic density matrix in the triple dot reads,
\begin{equation}
    \frac{d\hat{\rho}}{dt}
    = -i\big[\hat{H},\hat{\rho}\big] + \mathbf{\Gamma}\hat{\rho},
    \label{eq:tqd_von_neumann}
\end{equation}
with $\hat{H}=\hat{H}_0+\hat{H}_{\rm hf,mf}$.
We describe the transitions $(2,0,1)/(1,0,2)\to (1,0,1) \to (1,1,1)$ with the term $\mathbf{\Gamma}\hat \rho=-\frac{1}{2}\Gamma\{\hat{P}_{\rm dec},\hat{\rho}\}+\frac{1}{8}\Gamma(\mathbbm{1} - \hat P_{\rm dec})\rho_{\rm dec}$, where the operator $P_{\rm dec}=\sum_{i = D_{L,R}^\alpha} \ket{i}\bra{i}$ projects to the subspace that is coupled to the drain leads and $\rho_{\rm dec}=\sum_{i = D_{L,R}^\alpha} \rho_{i,i}$.

Assuming that $\Gamma$ is the largest energy scale involved, we again separate time scales and write the effective (1,1,1) Hamiltonian
\begin{equation}
    \hat{H}^{(1,1,1)} = \sum_{\alpha = \pm} -\alpha \frac{3}{2}E_{\rm Z} \ket{Q_{\alpha 3/2}}\bra{Q_{\alpha 3/2}} + \hat H^\alpha_{\frac{1}{2}},
\end{equation}
using the two $3\times3$ blocks
\begin{multline}
    \hat{H}_\frac{1}{2}^\alpha
    = -\alpha \frac{1}{2}E_Z  + 3E_D\ket{D_1^\alpha}\bra{D_1^\alpha} + E_D\ket{D_2^\alpha}\bra{D_2^\alpha}
    \\
    +\alpha \left(\begin{array}{ccc}
        0 &
        -\frac{\sqrt 2}{3}\Delta_M &
        \sqrt \frac{2}{3}\Delta_{LR}  \\
        -\frac{\sqrt{2}}{3}\Delta_M &
        -\frac{1}{3}\Delta_M &
        -\frac{1}{\sqrt{3}}\Delta_{LR} \\
        \sqrt\frac{2}{3}\Delta_{LR} &
        -\frac{1}{\sqrt{3}}\Delta_{LR} &
        \frac{1}{3}\Delta_M
    \end{array}\right),
\end{multline}
acting on the subspaces $\{\ket{Q_{\alpha 1/2}},\ket{D_{1}^\alpha},\ket{D_{2}^\alpha}\}$.
Here $E_{\rm Z}$ contains the contribution $\frac{1}{3}(K^z_L+K^z_C+K_R^z)$ from the average nuclear spin polarization.
We assumed $E_{\rm Z}$ to be large enough that we can neglect the transverse components $K_d^{x,y}$ that couple states with different $S^{\rm tot}_z$.
The projection to the (1,1,1) subspace introduced the exchange energy
\begin{align}
E_D=\frac{2t^2\epsilon_M}{4\epsilon_M^2+\Gamma^2},
\end{align}
and makes the states $\ket{D_1^\pm}$ and $\ket{D_2^\pm}$ decay with rates $\Gamma_1=3\Gamma_D$ and $\Gamma_2=\Gamma_D$, respectively, where 
\begin{align}
\Gamma_D=\frac{2t^2\Gamma}{4\epsilon_M^2+\Gamma^2}.
\end{align}

Assuming that the exchange energy $E_D$ is much larger than the gradients $\Delta_{LR}$ and $\Delta_M$, we diagonalize $\hat H^\pm_\frac{1}{2}$ using perturbation theory and thusly find expressions for the eigenstates and their decay rates valid to lowest order in the gradients~\footnote{Here we keep only the effect of $E_D$ on the structure of the basis, i.e., we disregard $\Gamma_D$.}.
For non-zero gradients, the occupation probabilities are approximately $\frac{1}{2}$ for $\ket{Q_{\pm 3/2}}$ and zero for the remaining six states. Like for the double dot, we then calculate the hyperfine-induced flip-flop rates perturbatively using Fermi's golden rule \eqref{eq:flip}, and translate the resulting flip rates to evolution equations for the average polarization $P_\Sigma = \frac{1}{3}\left(P_L+P_C+P_R\right)$, and the two polarization gradients $P_{LR} = \frac{1}{2}\left(P_L-P_R\right)$ and $P_M=P_C-\frac{1}{2}\left(P_L+P_R\right)$. This gives, to lowest order in the field gradients $\Delta_{LR}$ and $\Delta_M$
\begin{align}
\frac{dP_{LR}}{dt} = {} & {} \!-\!\left[G+\frac{1}{\tau}\right]P_{LR} - \frac{G}{E_D}\Delta_{LR},\label{eq:analytic_average_result_TQD}\\
\frac{dP_M}{dt} = {} & {} 
\!-\!\left[\frac{5}{3}G+\frac{1}{\tau}\right]P_{M}
-GP_\Sigma - \frac{2G}{3E_D}\Delta_M,\label{eq:analytic_Delta_result_TQD}\\
\frac{dP_\Sigma}{dt} = {} & {} \!-\!\left[\frac{4}{3}G+\frac{1}{\tau}\right]P_\Sigma - \frac{2}{9}GP_{M},\label{eq:analytic_DeltaM_result_TQD}
\end{align}
with $G=A^2\Gamma_D/4N^2E_{\rm Z}^2$, where we again assumed equal $N$ on all dots, for simplicity.

As in the double dot, all polarization gradients thus acquire an effectively enhanced relaxation rate.
We further find that the polarization dynamics of $P_{M}$ and $P_\Sigma$ are coupled, which is a result of the geometry of the source and drains.
However, for typical parameters the last terms in Eqs.~\eqref{eq:analytic_average_result_TQD} and \eqref{eq:analytic_Delta_result_TQD} dominate, predicting an efficient suppression of both gradients, similar to the double-dot case.

Using these results, we can again investigate the stochastic fluctuations around stable points, using a linear Fokker-Planck equation that describes the time-dependent probability distribution $\mathcal{P}(n,m,l,t)$, where $n=\frac{3}{2}N P_\Sigma$, $m=NP_{LR}$ and $l=\frac{2}{3}NP_{M}$.
In the continuous limit, and to lowest order in the gradients, we find a covariance matrix that reads
\begin{align}
\sigma_{LR}^2 = {} & {} \frac{1}{2N}\frac{2E_D}{2E_D+A},\label{eq:fluctuations_Delta_TQD}\\
\sigma_{M}^2 = {} & {} \frac{3}{2N}\frac{E_D(81E_D+10A)}{81E_D^2+27AE_D+2A^2},\label{eq:fluctuations_DeltaM_TQD}\\
\sigma_\Sigma^2 = {} & {} \frac{1}{3N}\left[1 - \frac{AE_D}{81E_D^2+27AE_D+2A^2}\right],\\
\sigma_{LR,M}^2 = {} & {} \frac{2}{N}\frac{AE_D}{81E_D^2+27AE_D+2A^2}\label{eq:slrm}.
\end{align}
Realistically $A \gg E_D$, so the r.m.s.\ of the fluctuations of the two gradients are suppressed by a factor $\sim \sqrt{E_D/A}$, whereas the fluctuations of $P_\Sigma$ are barely affected, similar to what we found for the double dot.

\subsection{Numerical simulations}

\begin{figure}[th!]
	\centering
	\includegraphics[width=0.47\textwidth]{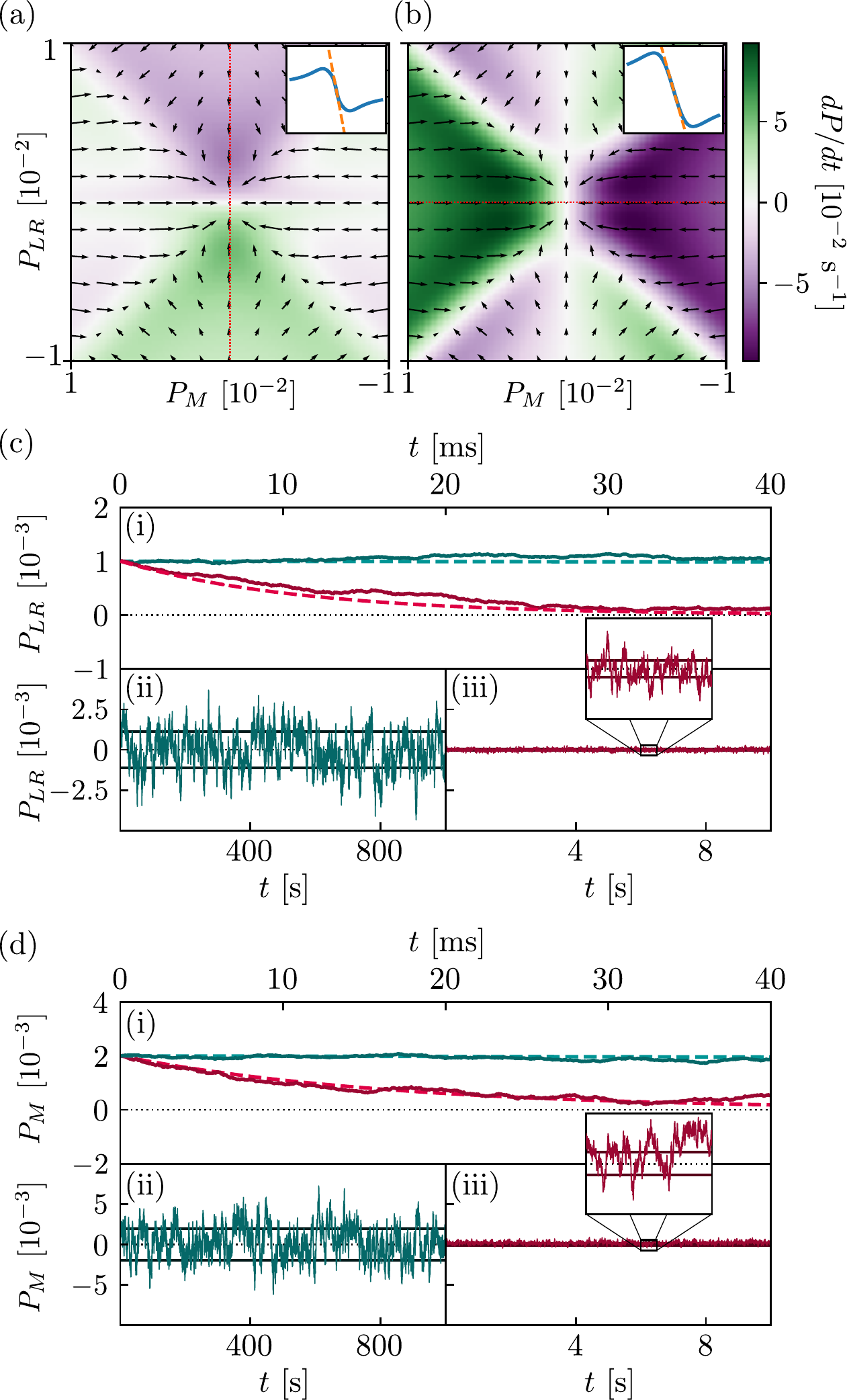}
	\caption{(a) $dP_{LR}/dt$ and (b) $dP_M/dt$ as a function of $P_{LR}$ and $P_M$ (color plots), calculated numerically using Eq.~(\ref{eq:flip_full}).
		The insets show line cuts along the red dashed lines, where the orange dashed lines indicate the slope of $dP/dt$ at the stable point as predicted by Eqs.~\eqref{eq:analytic_average_result_TQD} and \eqref{eq:analytic_Delta_result_TQD}.
		In both plots we also included the (same) vector field $(dP_M/dt, dP_{LR}/dt)$, represented by the black arrows. See the main text for all parameters used.
		(c,d) Simulated stochastic dynamics of (c) $P_{LR}$ and (d) $P_M$ with initial conditions $P_{LR}(0) = 0.001$, $P_M(0) = 0.002$, and $P_\Sigma = 0$.
		For the red lines we used the same parameters as in (a,b); the green lines show the dynamics in the absence of spin pumping (for $A=0$).
		Panels (i) show the short-time suppression toward zero gradients, where the dashed lines show the dynamics predicted by Eqs.~\eqref{eq:analytic_average_result_TQD}--\eqref{eq:analytic_DeltaM_result_TQD}.
		Panels (ii,iii) illustrate the fluctuations around the stable gradients at longer times.}
	\label{fig:pumping_triple_dot}
\end{figure}

Using the same method as in Sec.~\ref{sec:Numerical_simulation} we performed numerical simulations to corroborate our analytic results.
In Fig.~\ref{fig:pumping_triple_dot}(a,b) we first illustrate the coupled dynamics of $P_{LR}$ and $P_M$.
We set $P_\Sigma=0$, $A=125~\mu$eV, $E_Z=12.5~\mu$eV, $N=4\times10^5$, $\tau=5$~s, $\epsilon_M=100~\mu$eV, $\epsilon = 0$, $\Gamma=75~\mu$eV and $t=7.5~\mu$eV, and then we plot in color the rates of change $dP_{LR}/dt$ (a) and $dP_M/dt$ (b) as a function of $P_{LR}$ and $P_M$ as found using Eq.~(\ref{eq:flip_full}).
In both plots we also included the (same) vector field $(dP_M/dt, dP_{LR}/dt)$, represented by the black arrows, illustrating how both field gradients are indeed pumped toward zero.
The insets show line cuts along the red dotted lines, i.e., they show the rate of change of each polarization gradient as a function of the same gradient, where the other one is set to zero.
The dashed orange lines indicate the slope of the pumping curve at the stable point, as predicted by Eqs.~\eqref{eq:analytic_average_result_TQD}--\eqref{eq:analytic_DeltaM_result_TQD}, showing indeed good agreement with the numerical results.

In Fig.~\ref{fig:pumping_triple_dot}(c,d) we then show simulations of the stochastic dynamics of the two polarization gradients, performed in the same way as we did in Sec.~\ref{sec:Numerical_simulation} for the double dot.
We started with initial polarizations $P_{LR}(0) = 0.001$, $P_M(0) = 0.002$, and $P_\Sigma = 0$ and performed a simulation with the parameters given above (red lines) and one without spin pumping ($A=0$, green lines).
Panels (i) show the short-time dynamics, where the dashed lines correspond to the result predicted by Eqs.~(\ref{eq:analytic_average_result_TQD})--(\ref{eq:analytic_DeltaM_result_TQD}), and panels (ii) and (iii) show the long-time dynamics, where the horizontal solid lines indicate the r.m.s.\ value of the fluctuations as predicted from Eqs.~(\ref{eq:fluctuations_Delta_TQD})--(\ref{eq:slrm}).
We see that in all cases our analytic expressions agree reasonably well with the simulated dynamics of the gradients.
We further note that, for similar reasons as in the double dot, the average polarization $P_\Sigma$ drifts toward negative values, stabilizing around $\sim -0.004$.
Due to the way the dynamics of $P_M$ depend on $P_\Sigma$ [see Eq.~(\ref{eq:analytic_Delta_result_TQD})] one expects that the long-time stable polarization of $P_M$ is not at zero but at a small positive value; a careful look at Fig.~\ref{fig:pumping_triple_dot}(d,iii) shows that this is indeed the case in our simulations.

\subsection{Conclusion}
We found that electron transport through a linear triple quantum dot---with a source connected to the central dot and drains connected to the outer dots---tuned to the regime of Pauli spin blockade can yield a hyperfine-induced feedback cycle that dynamically suppresses the two nuclear polarization gradients in the triple dot.
To find the approximate magnitude of the r.m.s.\ value of the remaining nuclear-field fluctuations, we derived simple perturbative analytical expressions to describe the coupled dynamics of the polarization gradients.
This predicts a similar suppression of the fluctuations of the gradients as in the double-dot case, i.e., a suppression of one to two orders of magnitude.
We corroborated these analytic results with numerical simulations of the coupled electron-nuclear spin dynamics, finding good agreement between the two.

\section{Conclusion}\label{sec:conclusion}
In multielectron qubits, such as the double-dot-based two-electron singlet-triplet qubit and triple-dot-based three-electron exchange-only qubits, the main source of decoherence are usually the fluctuating nuclear-spin polarization gradients over neighboring dots.
These random gradients couple to the spins of the electrons in the dots and can thereby add to the qubit splitting or couple the two qubit states to each other as well as to other nearby states outside of the computational basis.

In this paper, we investigated the effect of running a DC current through such systems on the nuclear polarization gradients, while tuning to a regime of Pauli spin blockade.
We found that transport through the dots can give rise to a dynamical feedback cycle between the electronic and nuclear spins that results in an active suppression of the nuclear polarization gradients.

We considered a double-dot setup with and without significant spin-orbit interaction as well as a triple-dot setup without spin-orbit interaction.
For all cases we derived approximate analytical evolution equations for the nuclear polarization gradients, which all predict the possibility of a significant suppression of the fluctuations of the gradients.
We corroborated these results with numerical simulations of the stochastic coupled electron-nuclear spin dynamics which confirmed a reduction in the random fluctuations of the nuclear polarization gradients by one to two orders of magnitude.
These suppression mechanisms could thus present a straightforward way to significantly reduce the hyperfine-induced decoherence in multielectron qubits.

This work is part of FRIPRO-project 274853, which is funded by the Research Council of Norway (RCN), and was also partly supported by the Centers of Excellence funding scheme of the RCN, project number 262633, QuSpin.

%

\end{document}